\documentclass[12pt,preprint]{aastex}
\usepackage[backref,breaklinks,colorlinks,citecolor=blue]{hyperref}
\usepackage[all]{hypcap}

\def\nat{Nature}
\def\apj{ApJ}
\def\aj{AJ}
\def\apjl{ApJ}
\def\apjs{ApJ}

\def\araa{Ann. Rev. Astr. Ap}
\def\mnras{MNRAS}

\def\aap{A\&A}

\def\araa{ARA\&A}

\shortauthors{Cheng AND An}
\shorttitle{Parsec-scale radio structure of FR 0s}
\begin{document}

\title{Parsec-scale radio structure of 14 Fanaroff-Riley type 0 radio galaxies}

\author{X.-P. Cheng}
\affil{Shanghai Astronomical Observatory, Key Laboratory of Radio Astronomy, Chinese Academy of Sciences, Shanghai 200030, China}
\affil{University of Chinese Academy of Sciences, 19A Yuquanlu, Beijing 100049, China}

\author{T. An}
\affil{Shanghai Astronomical Observatory, Key Laboratory of Radio Astronomy, Chinese Academy of Sciences, Shanghai 200030, China; antao@shao.ac.cn}

\date{\today}

%------------------------------------------------------------------------------------------
%--- Abstract (ApJ max 250 words) -------------------------------------------------------------------------------
%------------------------------------------------------------------------------------------
\begin{abstract}

Recently a population of compact radio galaxies were classified as Fanaroff-Riley type 0 radio galaxies (FR 0s).
	The physical nature of FR 0s and the connection with the classical FR I and II galaxies are not currently well understood. Here, we report the radio properties of fourteen FR 0s on parsec (pc) scales derived from their very long baseline interferometry (VLBI) imaging observations.
	All sources show compact structures.
	Four sources show relativistic beaming with Doppler boosting factors ranging from 1.7 to 6.
	The brightness temperatures of the other ten are below the equilibrium limit.
	Jet proper motions are determined in two sources which have multiple epoch data, between $0.23\, c$ and $0.49\, c$, implying mildly relativistic jet flow.
	Low-amplitude flux density variation is found in J0943$+$3614 over a time period of 10 years. No significant variability are detected in three other sources over time scales of a few years.
	The radio properties of the FR 0s inferred from the VLBI data resemble GHz-peaked spectrum or compact steep-spectrum sources. Moreover, the diversity of their relativistic beaming indicators (brightness temperature, variability, jet proper motion) also imply that FR 0s might not be a homogeneous population of radio sources. 
	Detailed studies of the low-power ($P_{\rm 1.4GHz}<10^{24}$ W Hz$^{-1}$) FR 0 sources in the local Universe additionally offer a promising opportunity to understand their connection to the FR Is.

\end{abstract}


%------------------------------------------------------------------------------------------
%--- Introduction -----------------------------------------------------------------------
%------------------------------------------------------------------------------------------
\section{Introduction}

Radio galaxies were first observed more than six decades ago \citep[e.g.,][]{1953Natur.172..996J,1954ApJ...119..206B}. Studying their powerful jets can clarify mechanism of accretion and jet production in the vicinity of supermassive black holes and their large scale properties can help in understanding the connection with the host galaxy and its evolution.
Radio galaxies have been divided into two classes,  Fanaroff-Riley (FR) type I and II, based on their distinctive morphological differences \citep{1974MNRAS.167P..31F}. The FR II galaxies are characterized by compact, bright hotspots at the tips of the terminal lobes, weak or invisible core and inner jets; while the FR Is have prominent jets, midway hotspots and complex morphologies ranging from well confined lobes to extended plumes or tails, diffuse and amorphous features beyond hotspots \citep[e.g. in Cen A,][]{2009MNRAS.395.1999C}.
Moreover, FR IIs are clearly divided from FR Is with the FR IIs being typically much brighter in radio \citep[e.g.,][]{1996AJ....112....9L,2017MNRAS.466.4346M}.
Studies of jet evolution and its interaction with the host galaxy environment are generally focussed on the brighter FR IIs \citep[e.g.,][]{1997MNRAS.286..215K,2002MNRAS.336..649K,2007MNRAS.381.1548K,2012ApJ...760...77A}, though, the inferable physical properties of the jet (e.g., kinematics, thermodynamics and radiation), surrounding ISM and the accretion process have additionally been explored by modeling the more morphologically complex FR Is \citep[e.g.,][]{1995ApJS..101...29B,2002MNRAS.336.1161L,2009MNRAS.397.1113W,2010ApJ...713..398L,2015ApJ...806...59T}.

Recently \citet{2015A&A...576A..38B} observed a new class of compact radio galaxies with source size $<$5 kilo-parsec (kpc), referred to as FR 0s \citep{2009A&A...508..603B}. Currently, studied FR 0s in the radio are still limited to the available data mainly from the Karl G. Jansky Very Large Array (VLA) \citep{2018A&A...609A...1B} and the Australia Telescope Compact Array \citep[ATCA,][]{2014MNRAS.438..796S}. High-resolution images of FR 0s are essential possibly to further resolve compact emission structures and clarify their current classification as part of a continuous population of sources in parallel to the FR I/II as opposed to constituting their evolutionary phase. We attempt at addressing the above motivations by identifying and studying a sample of 14 FR 0s with available higher resolution radio images.

In this paper, we present the radio morphologies of the 14 FR 0 sources at sub-pc resolutions and report their core brightness, jet proper motion and variability.
Section 2 describes the sample selection and data reduction.
The VLBI observational results are presented in Section 3.
In Section 4, the radio classification of FR 0s and the possible evolutionary relationship with the large-sized FR Is are discussed.
The main results are then summarized in Section 5.
The cosmological parameters of H$_{0}$ = 73 km s$^{-1}$ Mpc$^{-1}$,  $\Omega_{\rm M}$ = 0.27 and $\Omega_{\Lambda}$ = 0.73 are adopted; 1 mas angular size corresponds to a projected linear size of 0.2 pc at $z = 0.01$.

\begin{figure}
  \centering
  \begin{tabular}{ccc}
  \includegraphics[width=0.3\textwidth]{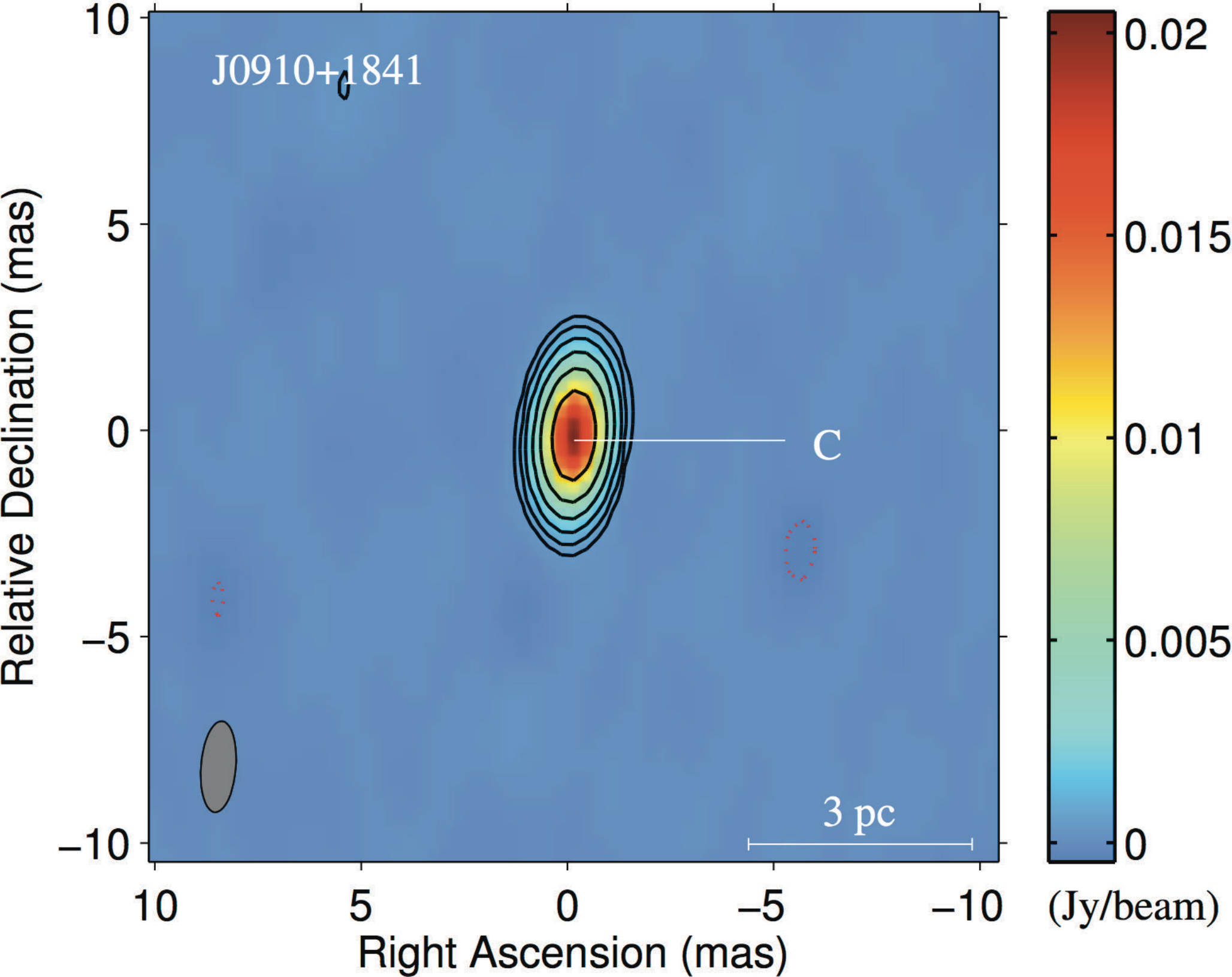} & \includegraphics[width=0.3\textwidth]{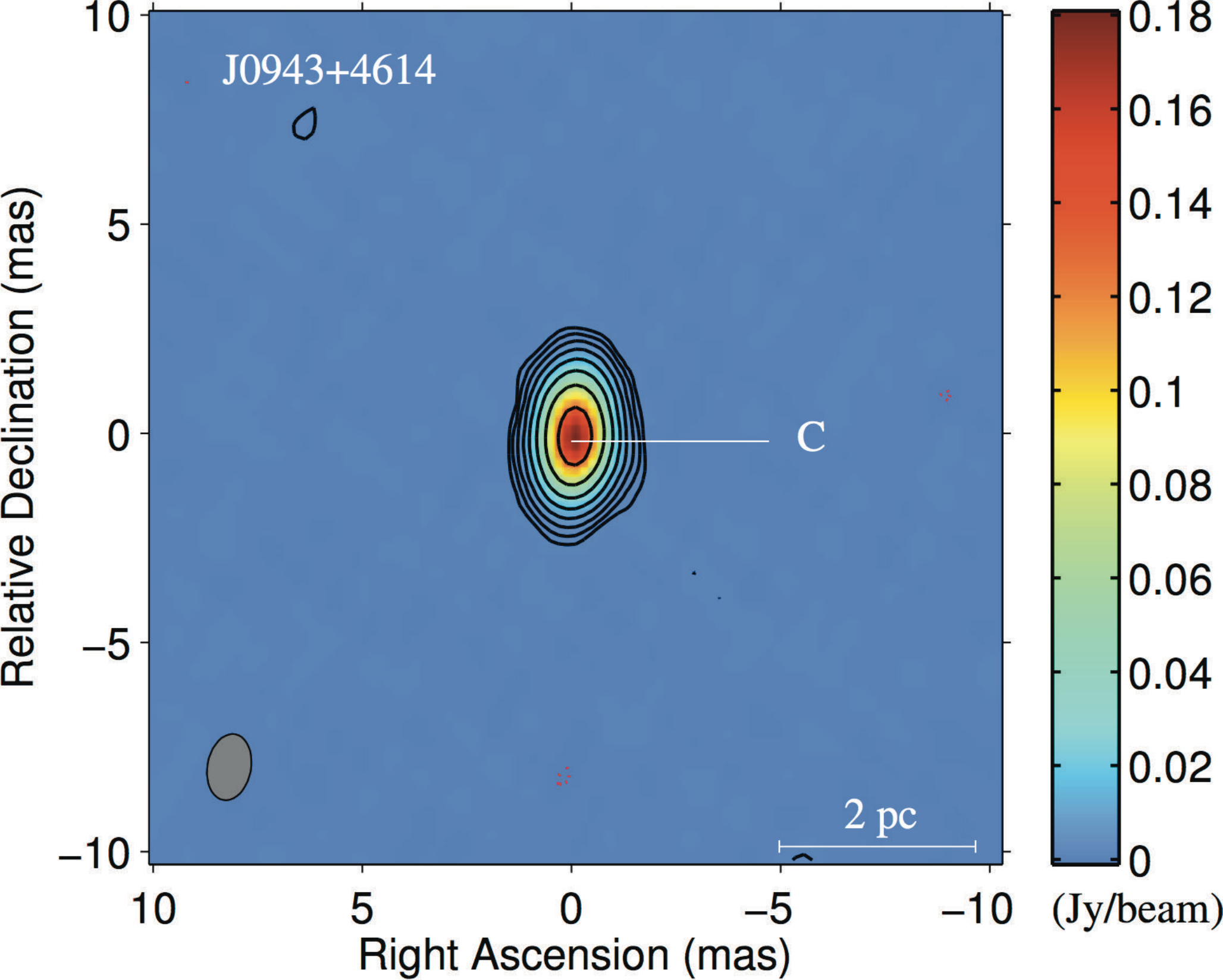}&\includegraphics[width=0.3\textwidth]{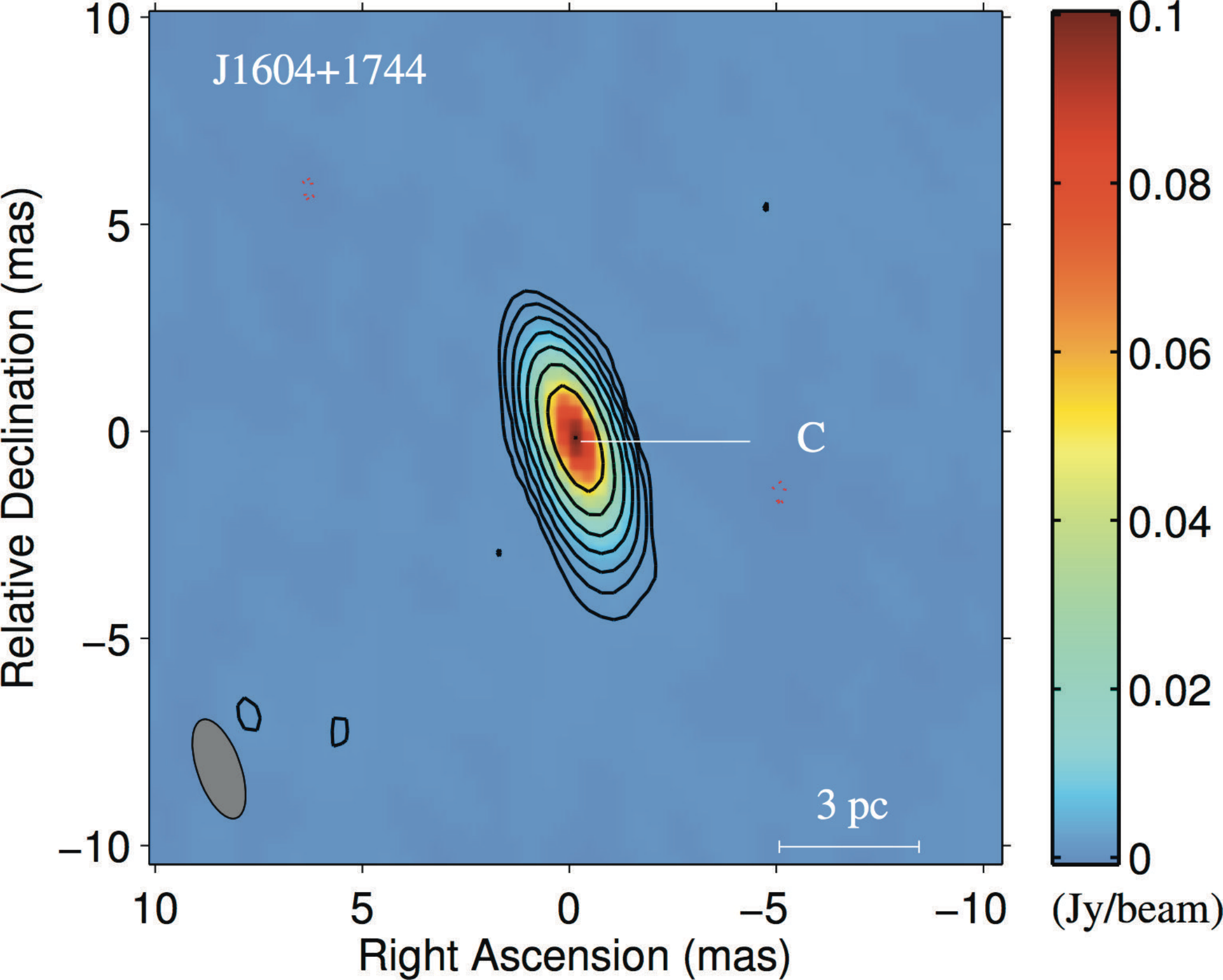} \\
  \includegraphics[width=0.3\textwidth]{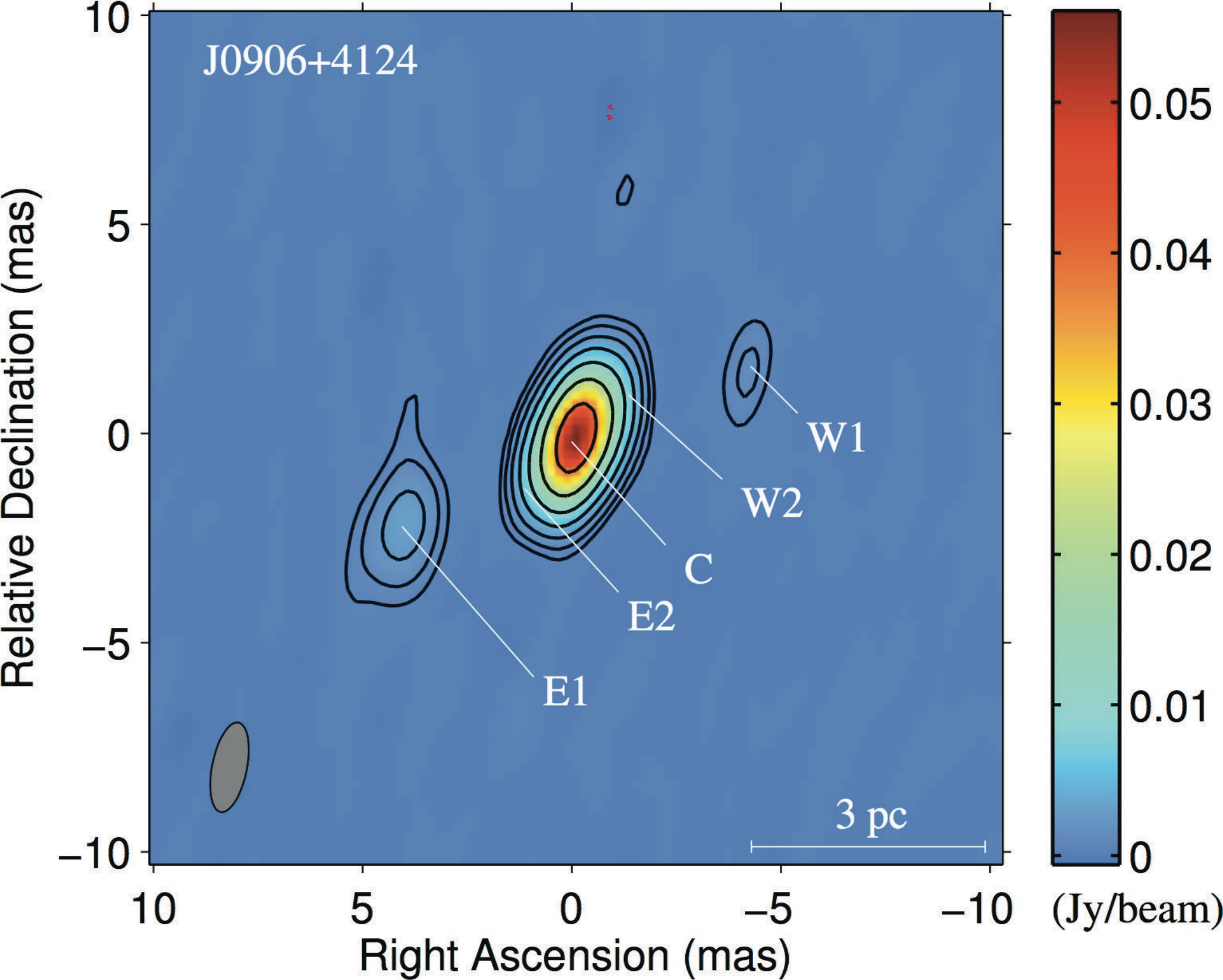} & \includegraphics[width=0.3\textwidth]{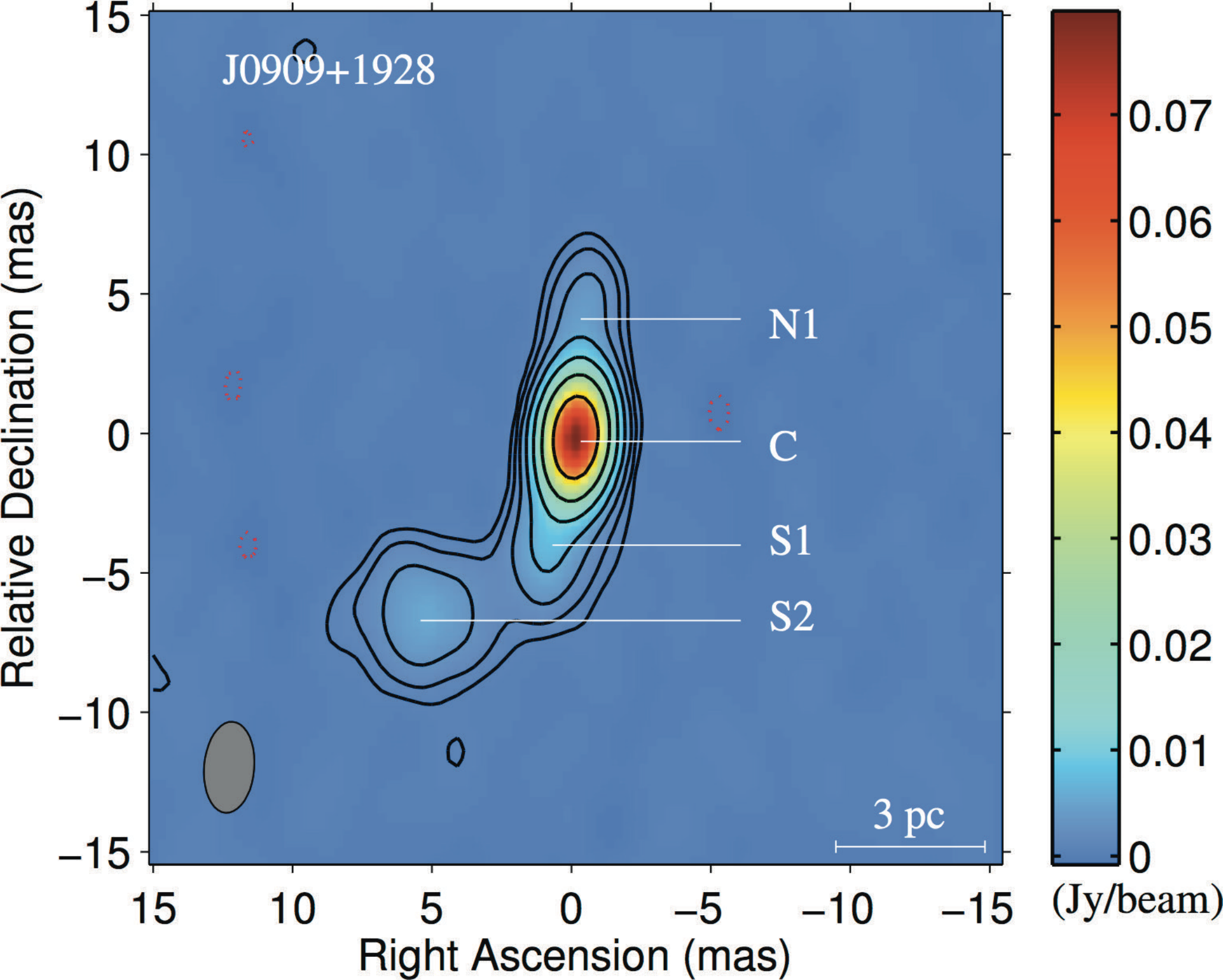}&\includegraphics[width=0.3\textwidth]{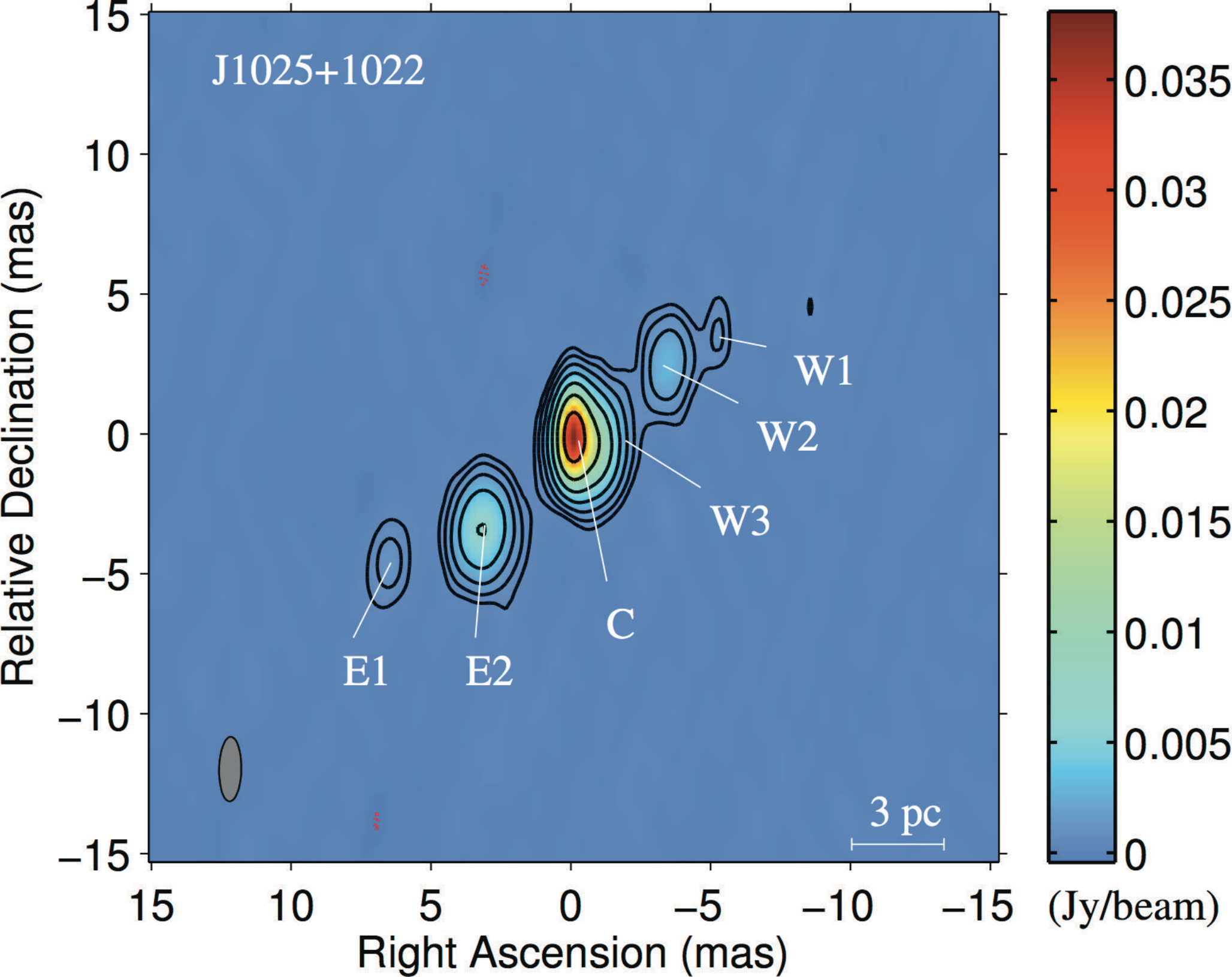} \\
  \includegraphics[width=0.3\textwidth]{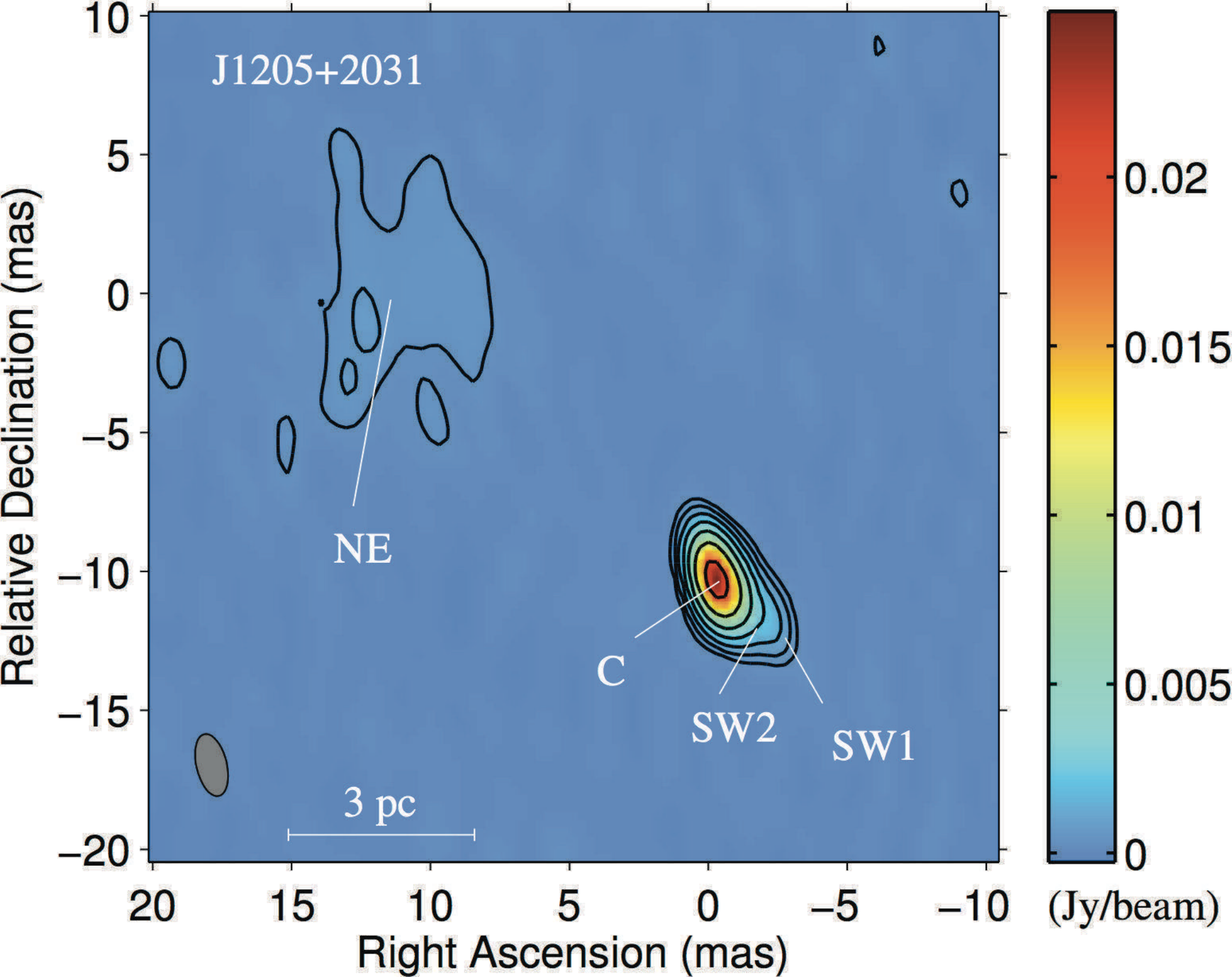} & \includegraphics[width=0.3\textwidth]{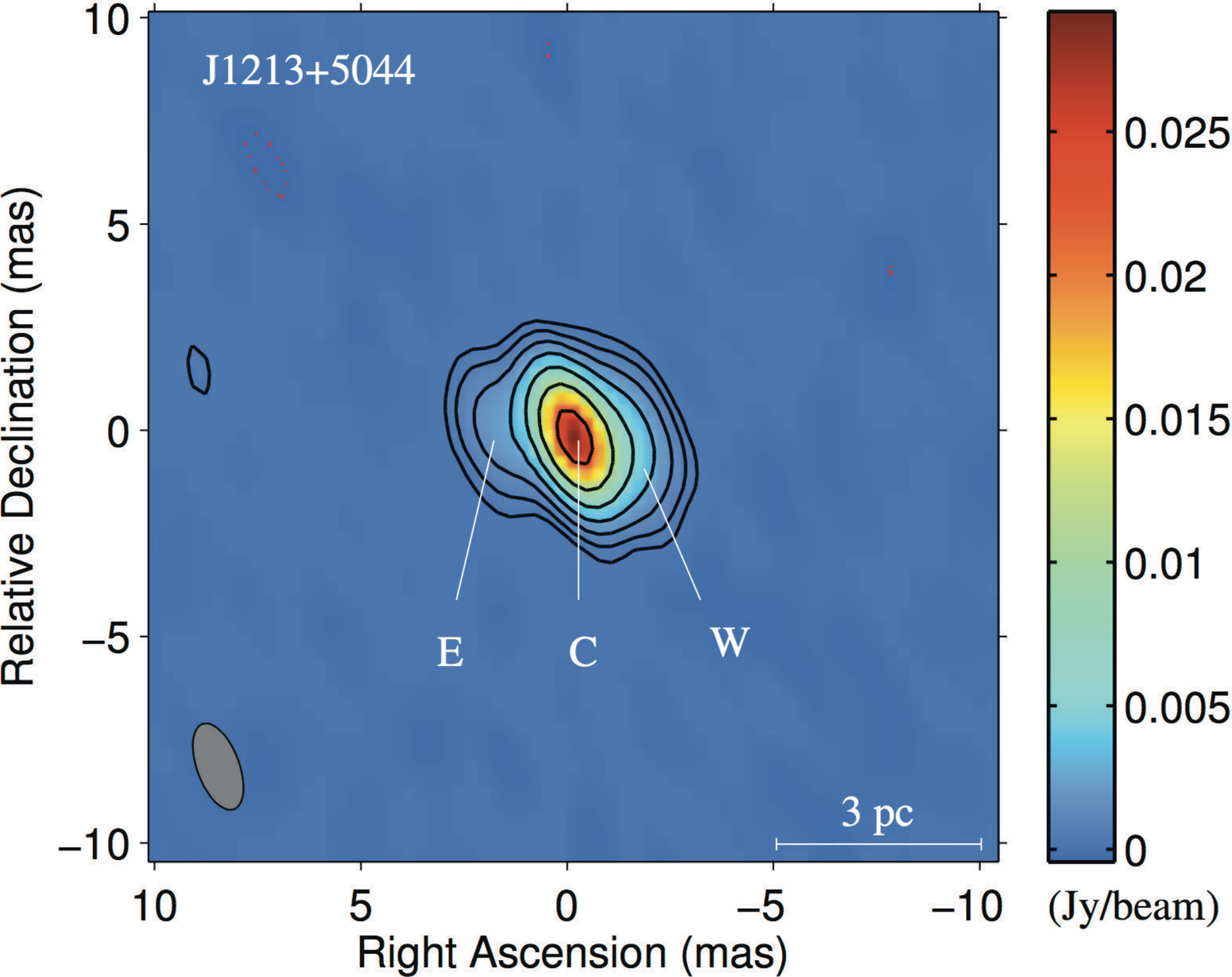}&\includegraphics[width=0.3\textwidth]{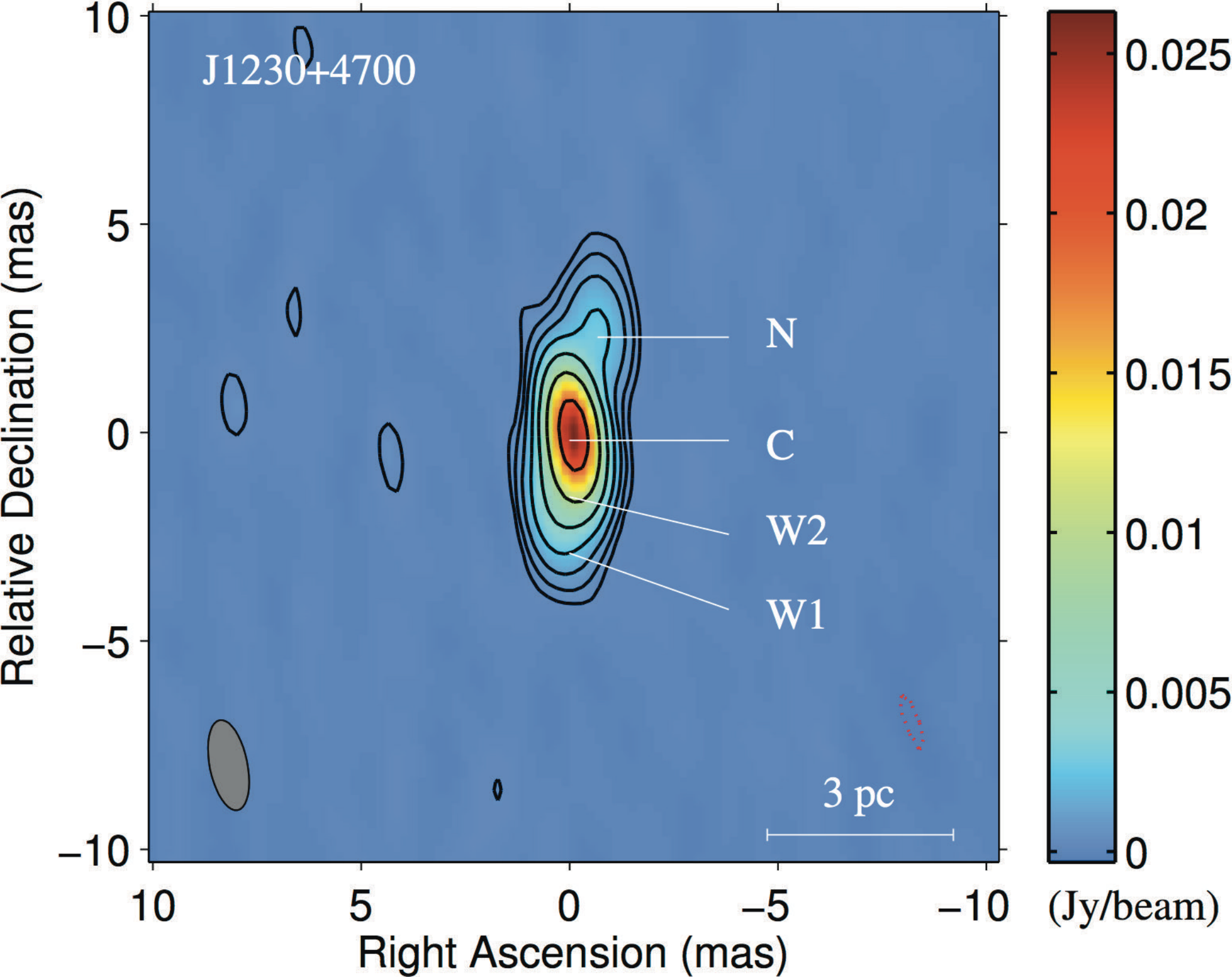}
  \\
      \end{tabular}
  \caption{Total intensity images of the sources showing compact cores, and core-twin jet structures. The images are made from 8.4 GHz VLBA data by using natural weighting.  The typical off-source {\it rms} noise is 0.2 mJy beam$^{-1}$. The lowest contour represents 3 times the {\it rms} noise.
  The contours are drawn at $-$1, 1, 2, 4, ..., $2^{\rm n}$ times the lowest contour level. The image parameters are referred to Table \ref{obs}.
 }
  \label{1}
  \end{figure}

\section{Sample selection and VLBI data reduction}

The study of \citet{2018A&A...609A...1B} identified 108 compact radio sources as FR 0s from amongst a larger sample of 18286 sources \citep{2012MNRAS.421.1569B} by using  selection criteria including $z \leq  0.05$, projected source size $< 5$ kpc and radio flux density $S_{\rm 1.4GHz}^{\rm VLA} > 5$ mJy.
We cross-matched the FR 0 catalog of \citet{2018A&A...609A...1B} with the mJIVE-20 catalog \citep{2014AJ....147...14D} and astrogeo database\footnote{VLBA calibrator survey data base is maintained by Leonid Petrov, http://astrogeo.org/.} resulting in an overlap consisting of 14 FR 0s listed in Table \ref{gentable}.
In this sub-sample of 14 sources, the flux densities are in the range 50-396 mJy with a mean of 106.4$\pm$23.5 mJy and the projected sizes are $<$ 3.5 kpc. The original sample of 108 sources has flux densities in the range 6-396 mJy with a mean of 34.3$\pm$4.3 mJy and the projected sizes are $<$ 5 kpc, indicating that these 14 sources are more compact and brighter than the general population. These then resemble the FR Is with compact bright central emission from 10-30 kpc \citep[e.g.,][]{2017A&A...598A..49C}, thus motivating an investigation of a possible physical connection between FR 0s and FR Is, or any evolution between two populations. It must be noted that the remaining (weaker) sources in the FR 0 catalog may not have similar energetics or morphological characteristics as the 14 sources, thus indicating that the evolutionary connection inferred from this study may not be generally applicable to the full sample; this then warrants a systematic deep VLBI imaging of the large sample to help clarify the physical nature of FR 0s.

The visibility data, acquired from the VLBI surveys used for astrometric and geodetic studies, are observed in snapshot mode.
In general, each source is observed with a few 5-minute scans within an observation session.
Table \ref{obs} gives a summary of the VLBI data.
Five sources (J0933$+$1009, J0943$+$3614, J1559$+$2556, J1604$+$1744, J1606$+$1814) have multiple-epoch datasets, allowing us to investigate their structure change and variability.
Four sources (J0933$+$1009, J0943$+$3614, J1559$+$2556, J1606$+$1814) have been observed simultaneously at dual frequencies, enabling the calculation of the spectral index that can be used to identify the flat-spectrum radio core.

The primary data reduction procedure, including flagging bad data points, amplitude calibration, ionosphere correction and fringe fitting, were performed using the Astronomical Imaging Processing Software ({\tt AIPS}) package of the National Astronomical Radio Observatory, the USA \citep{2003ASSL..285..109G}. We imported the calibrated VLBI data in {\tt Difmap} software package \citep{1997ASPC..125...77S} to carry out a few iterations of self-calibration in order  to eliminate the residual amplitude and phase errors. After self-calibration, the {\it u-v} visibility data were used to produce images.
A number of circular Gaussian components were fitted to the visibility data by using the {\tt MODELFIT} program  in {\tt Difmap} to quantitatively describe the emission structure.
The model fitting parameters are presented in Table \ref{model}.
The typical flux density uncertainty is about 10\%.
The uncertainties of the fitted component size is less than 1/5 times the beam size.
The separation of individual Gaussian components from the core is consistent within 15\% of the fitted size of the Gaussian models.

\begin{figure}
  \centering
  \begin{tabular}{ccc}
  \includegraphics[width=0.3\textwidth]{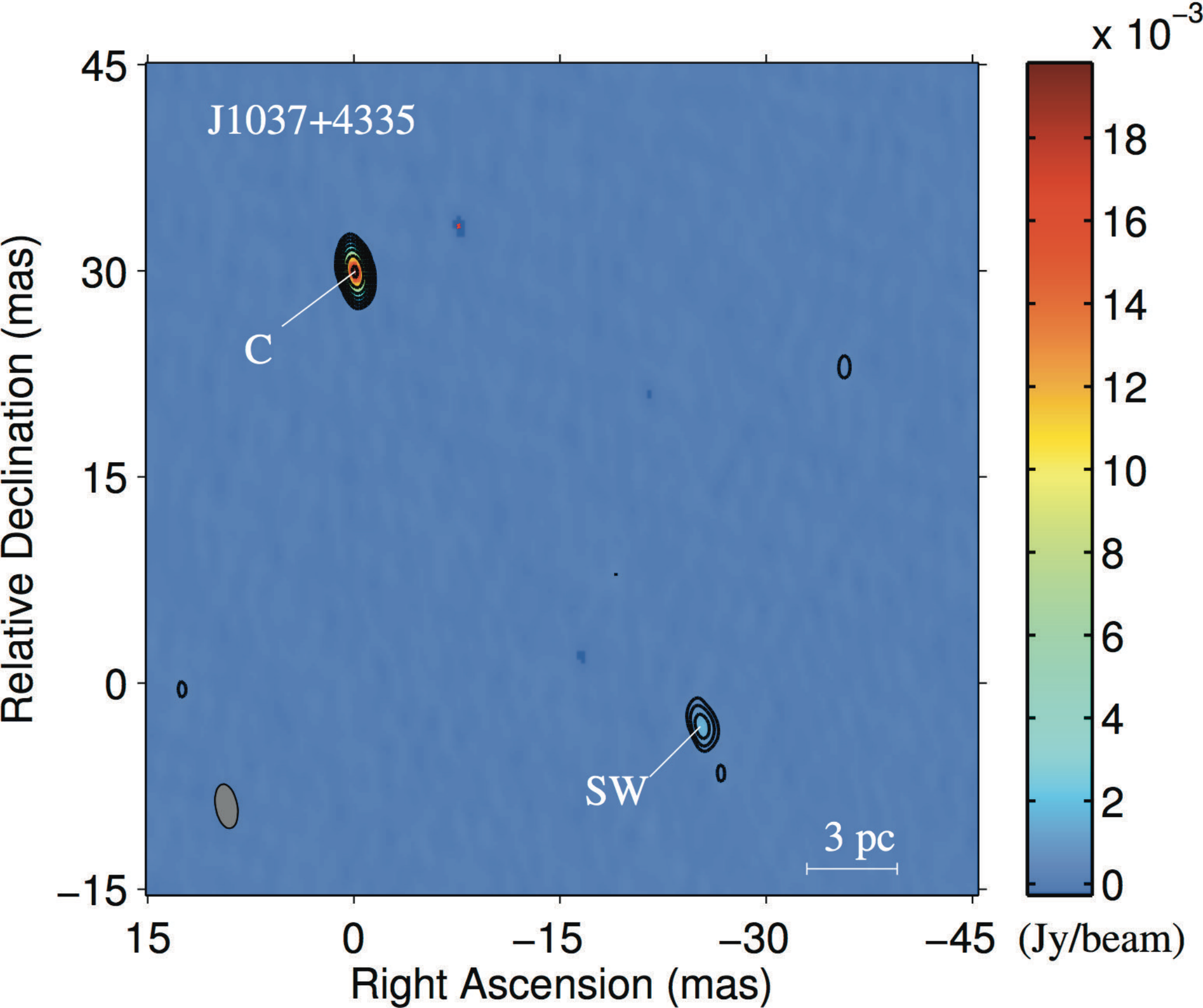} & \includegraphics[width=0.3\textwidth]{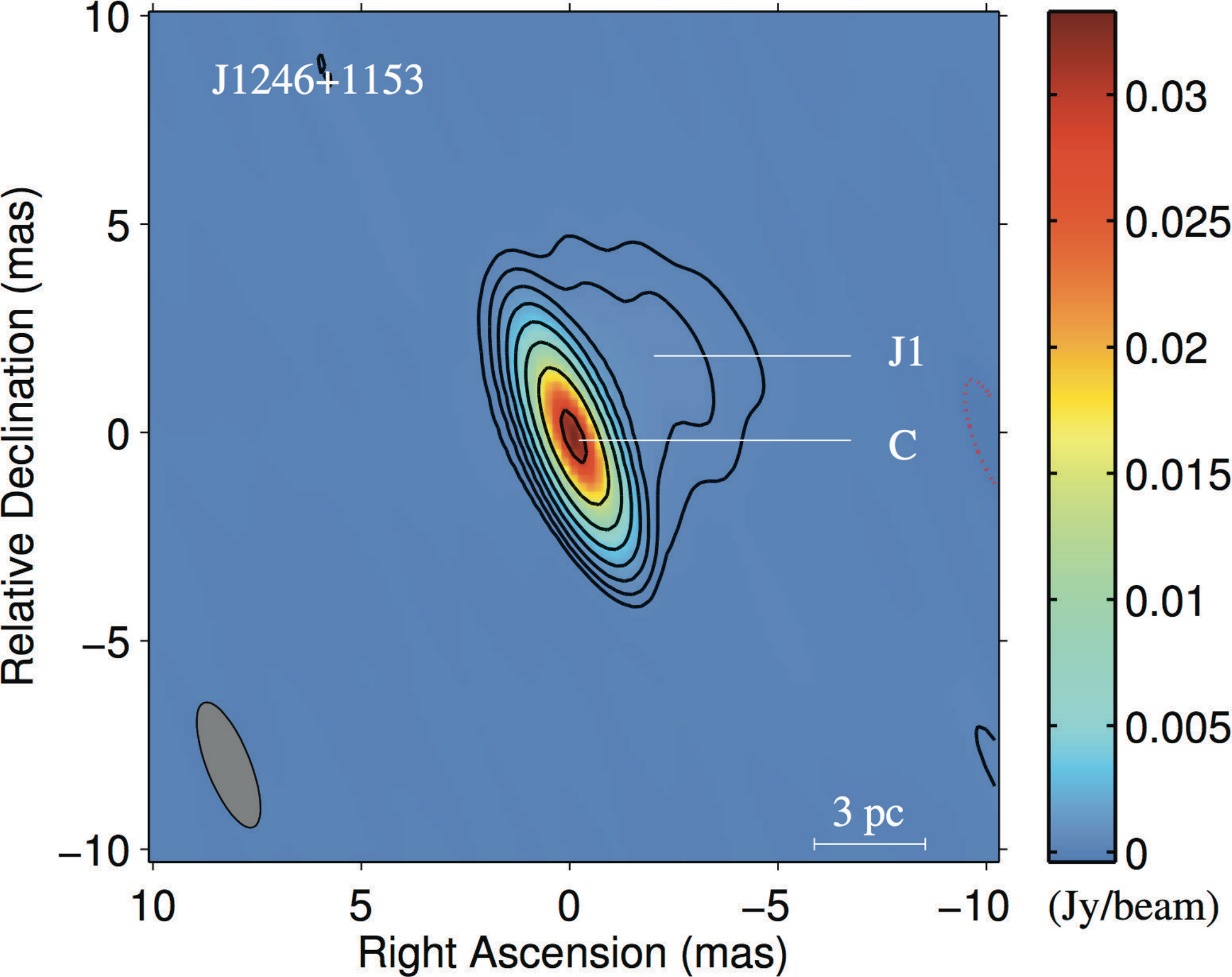} & \includegraphics[width=0.3\textwidth]{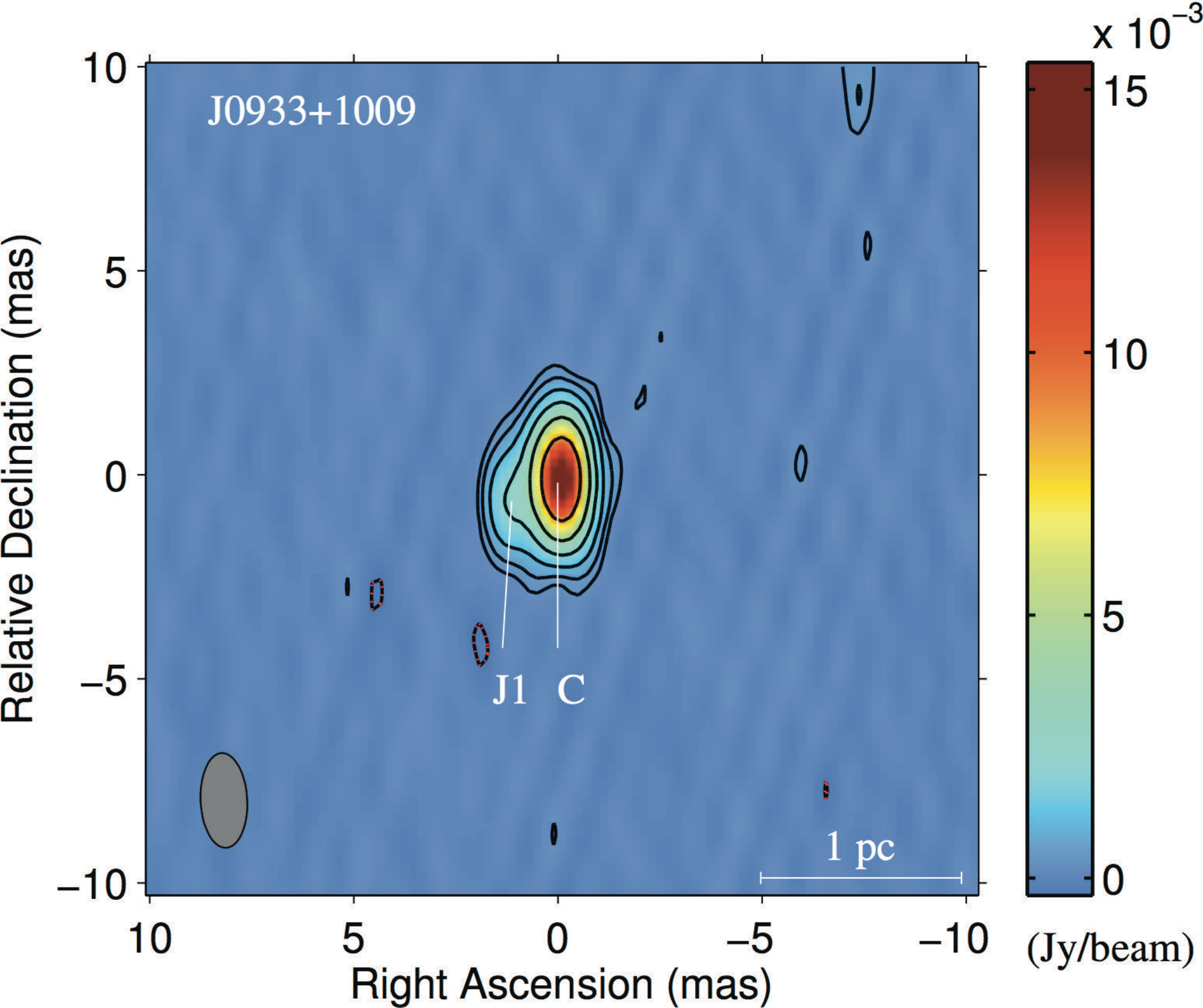} \\
  \includegraphics[width=0.3\textwidth]{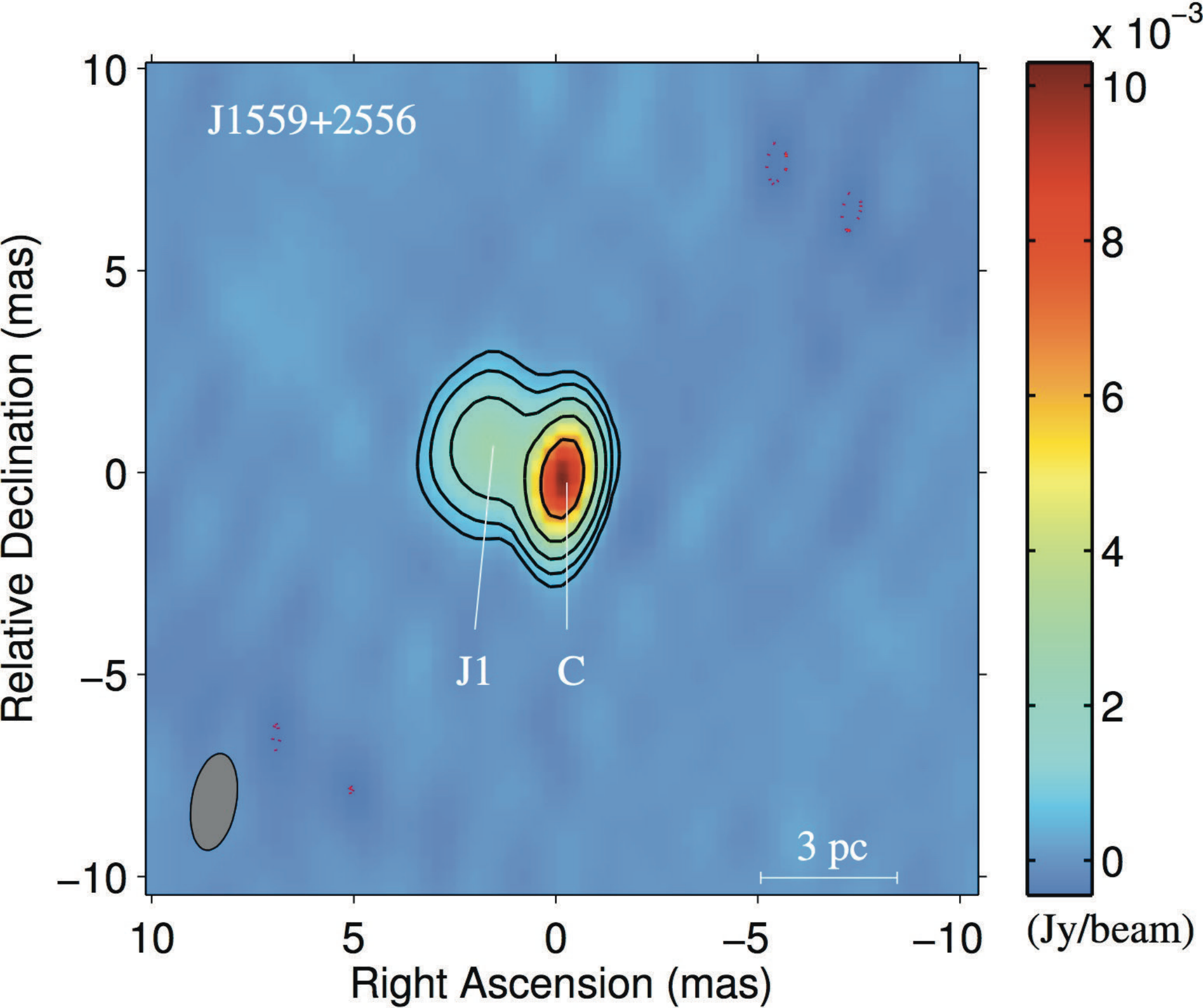} & \includegraphics[width=0.3\textwidth]{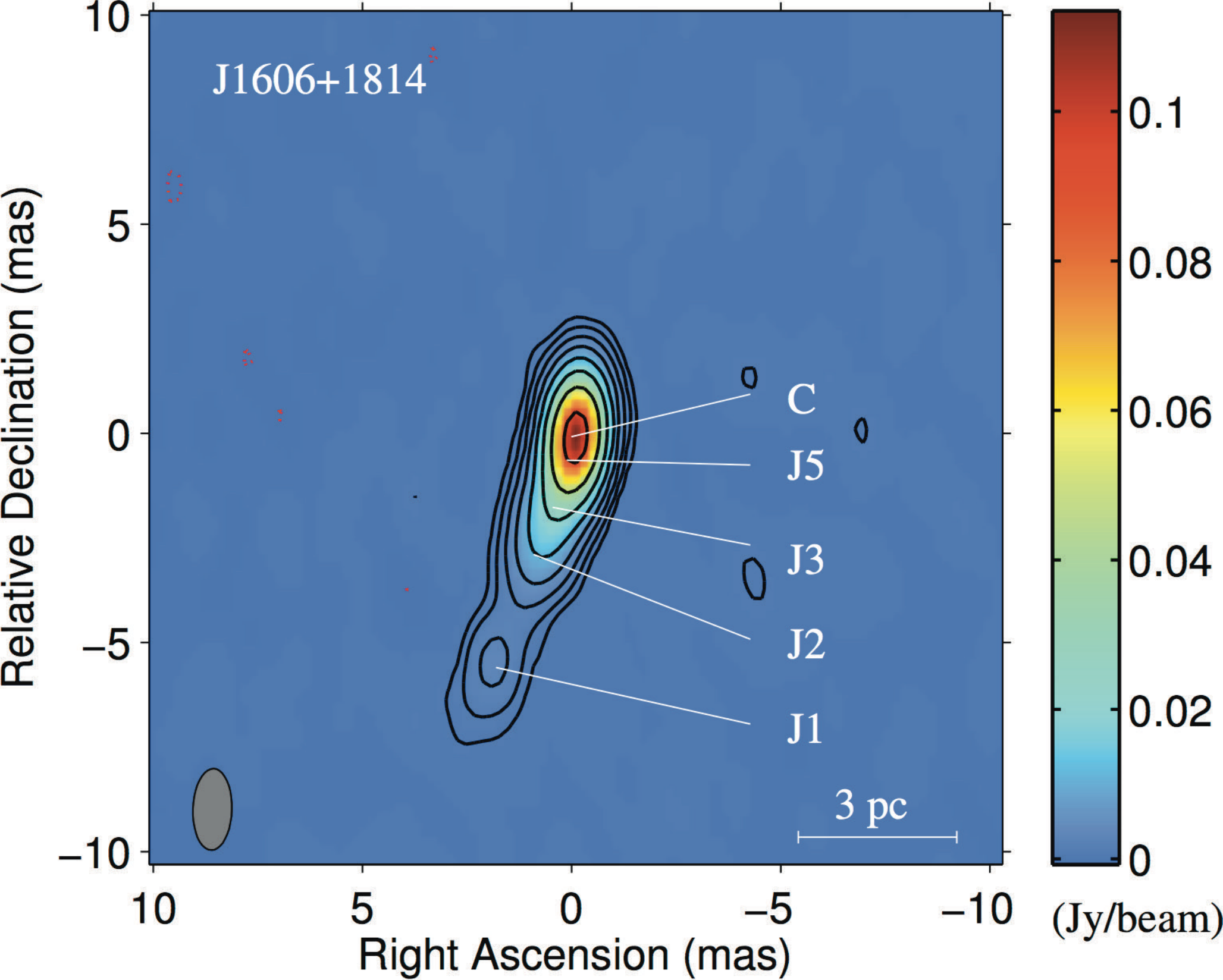} \\
  \end{tabular}
  \caption{Total intensity images of five sources showing core-jet structure. The images were derived from the 8.4 GHz VLBA data. The contours increase in step of two.
  }
  \label{2}
  \end{figure}

\section{Results}
\label{sec3}

\subsection{Pc-scale radio structure}

Figures \ref{1}--\ref{2} show the total intensity images of the 14 FR 0s.
For sources having multiple frequency data, we only display the highest-frequency 8.4 GHz images; for sources having multiple epoch data, the image on the latest epoch is shown.
The 14 sources display three types of morphologies:
(1) compact core; three sources (J0910$+$1841, J0943$+$3614 and J1604$+$1744) in Fig. \ref{1} show a single unresolved component, which is identified as the core;
(2) core and two-sided jets; the remaining six sources in Fig. \ref{1} show a bright central component straddled between two weaker symmetric features. The central components are the most compact and brightest, therefore we identify them as the core.
The compact triple morphology is analogous to compact symmetric objects (CSOs).
A diffuse feature is detected to the northeast of the core in J1205$+$2031; additional observations are required to clarify its physical nature as the counterjet or an artifact.
J0909$+$1928 displays a sharp jet bending at $\sim$5 mas (2.7 pc) south of the core.
The jets and counterjets in J0906$+$4124, J1025$+$1022, J1213$+$5044 and J1230$+$4700 are well aligned along the same direction.
(3) core and one-sided jet;
four sources (J0933$+$1009, J1246$+$1153,  J1559$+$2556, and J1606$+$1814) in Fig. \ref{2} show a resolved coherent structure with a compact component appearing at one end (identified as the core)
and contributing the most emission, representing a typical core-jet structure.
Indeed the bright, compact component in J0943$+$3614 and J1606$+$1814 have flat spectrum with spectral indices $\alpha$, 0.13 and $-$0.13. The jet extends to a maximal distance of $\sim$6 pc in J1606$+$1814. J1037$+$4335 is resolved into two isolated components, separated by 19.7 pc. The emission between the two components is fully resolved. The northeast component has a brightness temperature of $6 \times 10^{9}$ K (see Section 3.2), so we tentatively identify it as the core. Therefore the morphology of J1037$+$4335 is also identified as a core and one-sided jet; new VLBI observations at a different frequency confirm it as a flat spectrum core are needed.

Four sources (J0933+1009, J0943+3614, J1559+2556, J1606+1814) have simultaneous dual frequency VLBI data.
J0933+1009 and J1559+2556, with core-jet morphology, show steep spectral indices, $\alpha = -1.28$ and $\alpha = -1.12$, respectively.
J0943+3614 (unresolved core) and J1606+1814 (core-jet morphology) show flat spectral indices, $\alpha = 0.13$ and $\alpha = -0.13$, respectively.
The VLA observations of FR 0s by \citet{2015A&A...576A..38B} also present a diverse distribution of the spectral indices between $-1.16$ to $-0.04$, indicating both steep and flat spectra.
This implies that FR 0s as a class of sources are not homogeneous in their properties; they could be composed of sources ranging from the putatively youngest GHz-Peaked-Spectrum (GPS) to the evolved compact steep spectrum (CSS) sources \citep{2014MNRAS.438..796S,2015A&A...576A..38B}.

The observing frequency of the VLA data (1.4 GHz) is different from that of the VLBI data at 2.3, 5 and 8.4 GHz.
The spectral index is inferred first from the VLBI data (Column 4 in Table \ref{ratio of VLBI flux}), and then extrapolated to obtain the 1.4 GHz VLBI flux densities, which are listed in Column 3 in Table \ref{ratio of VLBI flux}.
The VLBI/VLA flux density ratios are presented in Column 5.
J0933$+$1009 and J0943$+$3614 have a ratio larger than 1.
The higher VLBI flux density could be due to variability with a variation factor higher than 3.
These seem erratic since FR 0s tend to show slow and low-amplitude flux density variation (Section 3.4).
Another possibility is that these two sources have a spectrum turnover between 1.4 and 2.3 GHz, similar with GPS sources, resulting in that the extrapolated 1.4 GHz VLBI flux density is overestimated.
This can be confirmed with future single-dish monitoring or simultaneous multi-frequency VLBI observations.
J1559$+$2556 and J1606$+$1814 show a higher VLA flux density than VLBI, indicating that a fraction of the emission is resolved with the VLBI.
Sub-arcsecond resolution images, e.g. obtained with the electronic Multi-Element
Remotely Linked Interferometer Network (e-MERLIN), are highly desirable to reveal the complete jet structure from pc to kpc scales.

\subsection{Core brightness temperature}

In all sources, the core dominates the total VLBI flux density.
We should note that, this could be attributed to selection effect because the VLBI survey sample comprises sources with flux densities normally higher than 50 mJy.
The core brightness temperatures based on the core flux densities and sizes, can be used as measure of the Doppler boosting and are listed in Table \ref{model}.
Doppler boosting is conventionally associated with relativistic jet seen at a small viewing angle.
The flat-spectrum core of J1606+1814 has the highest brightness temperature of $2.1 \times 10^{11}$ K. A Doppler boosting factor of 4.2 is inferred  from the ratio of $T_{\rm b,obs}/T_{\rm eq}$, where $T_{\rm eq} = 5 \times 10^{10}$ K is the equipartition brightness temperature  \citep{1994ApJ...426...51R}, indicating relativistic beaming of the jet.
J0933+1009 and J1559+2556 have brightness temperature of  $1\times 10^{10}$ K, and $2 \times 10^{10}$ K, respectively, below the equipartition limit.
Higher-frequency higher-resolution VLBI images may further constraint their $T_{\rm b}$s.
Two sources J0909+1928 and J1205+2031 show a core-twin-jet structure, similar to CSOs.
Their core brightness temperatures are higher than the equipartition limit.
The high $T_{\rm b}$ of the core and the morphology of symmetric outer jets could imply that the jet is initially beamed and has a sharp bending taking place within $\sim$0.2 pc.

\subsection{Jet proper motion}

Most sources have only one epoch data.
Among the five sources having multiple epoch data, two sources (J0943$+$3614, J1604$+$1744) are unresolved at 8.4 GHz and thus excluded.
The proper motions is then determined for the remaining three sources. Additional VLBI observations of a larger sample is essential to study the jet kinematics of FR 0s in a statistical manner.

The jet component in J0933$+$1009, at a distance of $\sim$1.4 mas from the core,
was not found to change in position significantly from 2012 January to 2015 September;
the resulting proper motion is $-0.04 \pm 0.07 \,c$ from the two-epoch data.
Detection such slow motions require a longer time span.
The jet component J1 in J1559$+$2556 displays a significant position change of $\sim$0.44 mas over a time separation of 2.6 yr, corresponding to a proper motion of $ 0.49 \pm 0.06 \,c$.
A total of four jet components are detected in J1606$+$1814. J4 appears in the late two epochs; since it is very close to the core, the model fitting of J4 is much affected by the mixture with the core emission and is hence not included in the calculation. The jet speeds of J1, J2 and J3 are $ 0.17 \pm 0.09 \,c$, $ 0.16 \pm 0.11 \,c$ and $ 0.23 \pm 0.08 \,c$,  respectively.
The jet speeds derived from these three FR 0s
show diverse distribution, ranging from moderate relativistic speed to stationary motion.

\subsection{Variability}

One of the 14 sources (J0943$+$3614) is monitored at 15 GHz with the OVRO 40 m telescope \citep{2011ApJS..194...29R} at intervals of two weeks. The data collection begins from 2008 August 14 until the latest epoch of 2018 February 2. From 2008 to 2014, the flux density remains quite stable. After 2014, the source becomes brighter, with the flux density increasing from $\sim$0.17 Jy in 2013 August to 0.25 Jy in 2017 April. The maximum variability amplitude is 40\%.
The compact source structure and {\bf slow} variability of J0943$+$3614 make it an ideal flux density calibrator in VLBI experiments.
Three sources (J0933+1009, J1559+2556, J1606+1814) have multi-epoch 8.4 GHz VLBI data, and we derived their variability factors of $<10\%$ for J0933+1009, $<20\%$ for J1559+2556 and $<15\%$ for J1606+1814 on time scales of several years.
These results are in agreement with \citet{2014MNRAS.438..796S} that the majority of FR 0s show low-amplitude variability in radio.

\section{Physical nature of FR 0s and their connection with large-scale FR Is}
\label{sec4}

The powerful large-scale FR I jets are commonly indicative of the jet interaction with local ISM which disrupt the jet morphology, and also shaping the properties of the intergalactic medium in galactic cluster environment \citep[AGN feedback,][]{2012ARA&A..50..455F}.
The recognition of similar features between the sub-sample of FR 0s studied here and the FR Is may address either their distinct natures or their evolutionary connection.

\begin{figure}[!htbp]
  \centering
  \begin{tabular}{c}
  \includegraphics[width=0.7\textwidth]{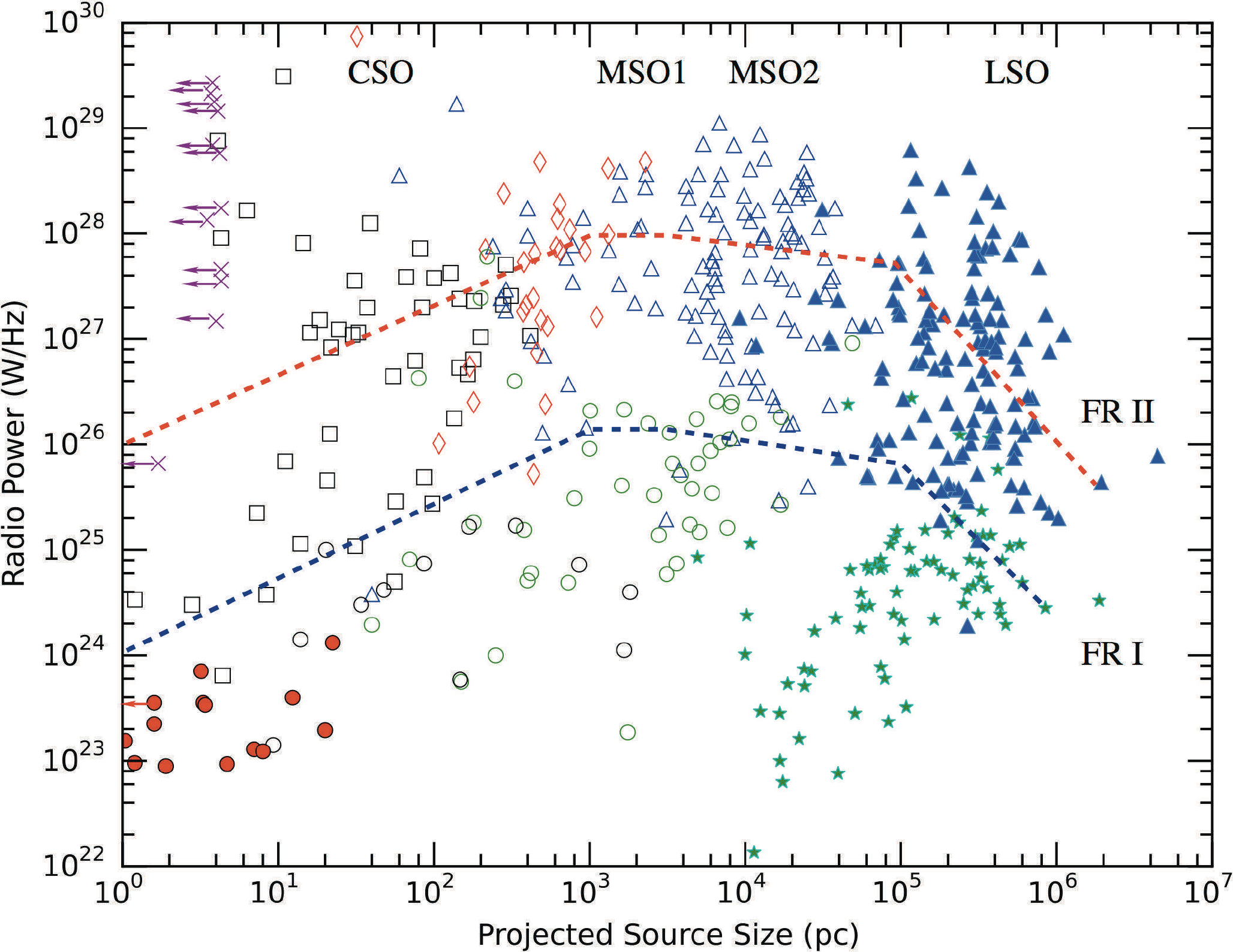} \\
    \end{tabular}
  \caption{Radio power versus source size of extragalactic radio sources. Red solid circles are FR 0 galaxies included in the present paper. red filled circle: FR 0s, black square: CSO, black circle: low-power GPS, red diamond: high-power GPS, purple cross: HFP, green circle: low-power CSS, blue open triangle: high-power CSS, blue filled triangle: FR II , and green filled star: FR I. The data are adopted from \citet{2012ApJ...760...77A}. Red and blue dashed lines are illustrative of the evolutionary tracks based on parametric modeling for the high power and low power sources, respectively. The FR 0 sources are situated in a bottom-left corner. This zone is occupied by low power CSOs and some compact low-power MSOs. }
  \label{P-D}
  \end{figure}

The dynamic evolution of radio galaxies can be represented at snapshots of radio power with the linear size of the radio structure.
The actual evolution is governed by the intrinsic physical properties of the central engine (the duration and persistence/intermittency of the nuclear activity, jet power), and environmental factors such as the density gradients of the ambient medium in the host galaxy, jet-ISM interactions along the path of the jets and within the lobes (see \citet{2012ApJ...760...77A} and references therein).
Figure \ref{P-D} shows the radio power P$_{ \rm 1.4GHz}$ versus the total extent of the source D (the so-called 'P-D' diagram).
The FR 0 galaxies studied here populate the bottom-left corner, marked in solid red circles.
This zone corresponds to the population of the low-power ($P_{ \rm 1.4GHz} <10^{24}$ W Hz$^{-1}$) CSOs in their earliest evolutionary phase.
Indeed, the pc-scale radio properties of some FR 0s inferred from the VLBI data in Section \ref{sec3}, including the compact symmetric morphology on pc scales, steep spectrum, moderate jet speed, slow structural and flux density variation, provide strong support to the association between FR 0s and low-power CSOs.
These 14 sources are the most compact ones and brightest in VLBI survey among the 108 FR 0s in \citet{2018A&A...609A...1B} with the remaining being even weaker at mJy level and hence, consistent with the above proposed scenario.
Some FR 0s may have large jet structure that is resolved in the VLBI images, therefore their positions in the P-D diagram may shift horizontally into the region of Medium-sized Symmetric Objects (MSOs).
In this sense, FR 0s could be a composite population of low-power CSOs and MSOs,
as is also manifested from the diverse distribution of their radio properties discussed in Section 3.
\citet{2018A&A...609A...1B} and \citet{2014MNRAS.438..796S} identified FR 0s as GPS and CSS sources based on the spectral index.
This is essentially consistent with our implication.

The dashed lines in Fig. \ref{P-D} denotes the evolutionary tracks based on parametric modeling.
These represent the expected evolutionary route of a radio source which begins as a CSO and successfully evolves to FR II/Is under conditions sustaining of long-duration AGN activity.
If this same evolutionary scenario is also applicable to the low-power FR 0 sources, this explains the formation of at least a part of FR Is.
The actual process of such evolution may be complicated due to the jet instability and jet power continuity.
According to \citet{2012ApJ...760...77A}, only a small fraction of low-power CSOs can develop to large structures.
This is as the characteristic distance determined by the jet power and the external medium density distribution is comparable to the projected separations inferred. As a result, the low power jets in these FR 0s can host boundary layers that rapidly transition from laminar to turbulent and in the process, significantly lose their energy and momentum.
Then, these FR 0s will end up as relics at scales of 0.1--1 kpc with halted evolution and can be targets for the next-generation high-sensitivity surveys, such as the Square Kilometre Array (SKA) continuum survey \citep{2015aska.confE.173K}.
If this scenario can be confirmed with further observations and supplemented by theoretical modeling, it provides a mechanism of the formation of some low-power FR Is.
Discovering progenitors of FR Is amongst FR 0s (or generally, the low-power CSOs) which were formed in the local Universe is of particular interest in studying the evolution of their host galaxies, some of which are spirals \citep{2014MNRAS.438..796S}.

\section{Summary}

The pc-scale radio properties of 14 FR 0s are investigated from their available VLBI higher resolution images. All sources show compact structure, either a single core, resolved core and single jet, or core and two-sided jets. In all sources, the core dominates the integrated VLBI flux density.
The properties derived from VLBI data show diverse distribution, indicating that FR 0s could be a mixed population comprising GPS and CSS sources, or a mixture of CSO and MSO in morphology nomenclature.
Comparing the extrapolated VLBI flux density with the 1.4 GHz VLA flux density suggests that at least two sources have a compact structure on both mas and arcsecond scales.
The core brightness temperatures of four sources are higher than the equipartition limit, indicating Doppler boosting.
Lower brightness temperatures in other sources indicate that the core emission is likely blended with that of the inner jet.
Additional new VLBI observations at higher frequencies and resolutions can place more stringent constraints on the core brightness temperatures.
Four sources exhibit stable flux densities and slow jet proper motions.
The derived VLBI properties of these FR 0s are consistent with them being low-power CSOs.
The fast approaching radio facilities, such as the SKA and the Jansky VLA, offer promising opportunities to discover a large population of FR 0s, enriching our understanding of the extragalactic radio source families and contributing towards presenting a complete picture of radio galaxy evolution.

\acknowledgments

We thank the referee for her/his helpful comments which improve the manuscript.
This work is partly supported by the SKA pre-research grant of the Ministry of Science and Technology of China (No. 2018YFA0404600) and the Chinese Academy of Sciences (CAS, No. 114231KYSB20170003).
We are grateful to helpful comments from Gabriele Giovannini, Elaine Sadler and Prashanth Mohan.
TA thanks the grant supported by the Youth Innovation Promotion Association of the CAS.
The National Radio Astronomy Observatory is a facility of the National Science Foundation operated under cooperative agreement by Associated Universities, Inc.

\label{lastpage}

%------------------------------------------------------------------------------------------
%--- Reference ------------------------------------------------------------------------------
%------------------------------------------------------------------------------------------

\appendix

\section{VLBI observations and model fitting parameters}

\begin{deluxetable}{ccccccc}
\centering
\tablecolumns{6}
\tabletypesize{\small}
\tablewidth{0pt}
\tablecaption{\label{gentable} Radio properties of the selected FR 0 galaxies.}
\tablehead{
\colhead{J2000}
& \colhead{$z$}
& \colhead{$S^{\rm VLA}_{\rm 1.4GHz}$}
& \colhead{$\rm L_{NVSS}$}
& \colhead{$P_{\rm 1.4 GHz}$}
& \colhead{$D_{\rm VLA}$}
& \colhead{Morphology}
\\
\colhead{Name}
& \colhead {}
& \colhead {(mJy)}
& \colhead {($ \rm erg s^{-1})$}
& \colhead{W Hz$^{-1}$}
& \colhead {(kpc)}
& \colhead {} \\
\colhead{(1)}   & \colhead{(2)}       & \colhead{(3)}            & \colhead{(4)}             & \colhead{(5)}                             & \colhead{(6)}                  & \colhead{(7)}              }
\startdata
J0906+4124 & 0.027 & 51.8   & 39.12 & 22.97 & 2.084 & two-sided Jet   \\
J0909+1928 & 0.028 & 69.1   & 39.26 & 23.11 & 2.156 & two-sided Jet   \\
J0910+1841 & 0.028 & 50.0   & 39.14 & 22.98 & 2.156 & point source    \\
J0933+1009 & 0.011 & 56.6   & 38.34 & 23.19 & 0.864 & One-sided Jet   \\
J0943+3614 & 0.022 & 75.1   & 39.10 & 22.95 & 1.708 & One-sided Jet   \\
J1025+1022 & 0.046 & 76.6   & 39.75 & 23.60 & 3.468 & two-sided Jet   \\
J1037+4335 & 0.025 & 132.2  & 39.44 & 23.29 & 1.932 & One-sided Jet   \\
J1205+2031 & 0.024 & 89.9   & 39.24 & 23.09 & 1.856 & two-sided Jet   \\
J1213+5044 & 0.031 & 96.5   & 39.50 & 23.35 & 2.380 & two-sided Jet   \\
J1230+4700 & 0.039 & 93.8   & 39.70 & 23.55 & 2.964 & two-sided Jet   \\
J1246+1153 & 0.047 & 61.2   & 39.68 & 23.53 & 3.540 & One-sided Jet   \\
J1559+2556 & 0.045 & 144.7  & 40.00 & 23.85 & 3.396 & One-sided Jet   \\
J1604+1744 & 0.041 & 96.0   & 39.75 & 23.55 & 3.112 & One-sided Jet   \\
J1606+1814 & 0.037 & 396.0  & 40.27 & 24.12 & 2.820 & One-sided Jet   \\
\enddata
\tablecomments{Columns are as follows:
(1) J2000 name;
(2) redshift;
(3) NVSS 1.4 GHz flux density from \citet{1998AJ....115.1693C};
(4) logarithm of the NVSS radio luminosity \citet{1998AJ....115.1693C};
(5) the source radio power;
(6) upper limit of the source size, assuming all the source is 4 arc min;
(7) morphology of the source;
}
\end{deluxetable}

\begin{deluxetable}{cccccccccc}
\centering
\tablecolumns{9}
\tabletypesize{\small}
\tablewidth{0pt}
\tablecaption{\label{obs} VLBI observation results}
\tablehead{
\colhead{J2000} & \colhead{Code} & \colhead{$\rm \nu$} & \colhead{Epoch}         & \colhead{$ \rm \tau$} & \colhead{BW} & \colhead{$B_{\rm maj}$} & \colhead{$B_{\rm min}$} & \colhead{$\theta$} & \colhead{$ \rm \sigma$}      \\
\colhead{(Name)}  & \colhead {}    & \colhead {(GHz)}    & \colhead {(yyyy-mm-dd)} & \colhead {(s)}        & \colhead {(MHz)}    & \colhead {(mas)}        & \colhead {(mas)}   &  \colhead{($\degr$)}   & \colhead {(mJy beam$^{-1}$)}     \\
\colhead{(1)}   & \colhead{(2)}  & \colhead{(3)}       & \colhead{(4)}           & \colhead{(5)}         & \colhead{(6)}       & \colhead{(7)}           & \colhead{(8)}           & \colhead{(9)} & \colhead{(10)} }
\startdata
J0906+4124   &   bc201a0  &  8.355  &  2102-04-22   &  629  &     128  &  2.17  &  0.83  &  $-$11.6  &  0.19  \\
J0909+1928   &   bt085b   &  4.845  &  2006-01-28   &  500  &     32   &  3.24  &  1.78  &  $-$4.69  &  0.26  \\
J0910+1841   &   bc201a2  &  8.355  &  2012-05-06   &  1280 &     128  &  2.34  &  1.11  &  $-$4.54  &  0.16  \\
J0933+1009   &   bc196zr  &  8.355  &  2012-01-13   &  320  &     128  &  2.38  &  0.96  &     7.86  &  0.20  \\
             &   bg129f   &  2.292  &  2014-12-20   &  300  &     128  &  7.39  &  3.79  &  $-$3.51  &  0.23  \\
             &   bg129f   &  8.668  &  2014-12-20   &  300  &     384  &  1.96  &  1.08  &  $-$3.85  &  0.17  \\
             &   bp192a8  &  7.624  &  2015-09-15   &  200  &     256  &  2.30  &  1.13  &     2.71  &  0.27  \\
             &   bp192a8  &  4.845  &  2015-09-15   &  214  &     256  &  3.99  &  1.63  &  $-$1.64  &  0.15  \\
J0943+3614   &   bk124    &  2.309  &  2005-07-09   &  270  &     32   &  8.44  &  3.72  &  $-$8.62  &  0.5   \\
             &   bk124b   &  8.646  &  2005-07-09   &  240  &     32   &  2.30  &  1.02  &  $-$8.03  &  0.51  \\
             &   bt085h   &  4.845  &  2006-05-01   &  600  &     32   &  3.53  &  2.27  &     20.1  &  0.23  \\
             &   bc201a2  &  8.355  &  2012-05-06   &  280  &     128  &  2.11  &  0.853 &  $-$2.23  &  0.27  \\
             &   bg219e   &  2.276  &  2014-08-09   &  160  &     256  &  6.09  &  4.05  &  $-$10.7  &  0.31  \\
             &   bg219e   &  8.355  &  2014-08-09   &  160  &     384  &  1.59  &  1.05  &  $-$11.6  &  0.28  \\
J1025+1022   &   bc196zs  &  8.335  &  2012-01-23   &  640  &     128  &  2.29  &  0.79  &  $-$1.07  &  0.17  \\
J1037+4335   &   bc196c   &  8.355  &  2011-01-20   &  960  &     128  &  2.16  &  1.05  &    10.40  &  0.09  \\
J1205+2031   &   bc201am  &  8.355  &  2012-03-07   &  640  &     128  &  2.26  &  1.08  &    13.40  &  0.14  \\
J1213+5044   &   bc201ak  &  8.355  &  2012-02-25   &  640  &     128  &  2.15  &  1.00  &    20.40  &  0.15  \\
J1230+4700   &   bc201am  &  8.355  &  2012-03-07   &  640  &     128  &  2.18  &  0.89  &    11.30  &  0.15  \\
J1246+1153   &   bc201a4  &  8.355  &  2012-06-02   &  320  &     128  &  3.18  &  1.03  &    21.71  &  0.12  \\
J1559+2556   &   bc191p   &  8.640  &  2010-12-12   &  480  &     128  &  2.91  &  1.43  &    33.24  &  0.22  \\
             &   bc196f   &  8.335  &  2011-02-07   &  480  &     128  &  2.08  &  0.84  &     8.69  &  0.18  \\
             &   bp171ae  &  4.845  &  2013-07-08   &  750  &     256  &  3.76  &  1.82  &  $-$5.98  &  0.07  \\
             &   bp171ae  &  7.624  &  2013-07-08   &  720  &     256  &  2.35  &  1.13  &  $-$9.29  &  0.17  \\
J1604+1744   &   bt085    &  4.845  &  2006-04-03   &  400  &     32   &  3.31  &  1.74  &  $-$4.28  &  0.21  \\
             &   bc196zo  &  8.355  &  2012-01-08   &  120  &     128  &  2.45  &  1.02  &    19.23  &  0.33  \\
J1606+1814   &   bk124b   &  8.646  &  2005-07-20   &  300  &     32   &  2.09  &  0.84  &  $-$5.40  &  0.69  \\
             &   bk124b   &  2.309  &  2005-07-20   &  260  &     32   &  8.91  &  3.15  &  $-$9.99  &  0.59  \\
             &   bt085    &  4.845  &  2006-04-03   &  600  &  32   &  3.54  &  2.31  &  $-$2.81  &  0.31  \\
             &   bc196zn  &  8.355  &  2011-12-19   &  180  &     128  &  2.81  &  1.27  &  $-$11.13 &  0.47  \\
             &   bg219g   &  2.292  &  2015-01-22   &   90  &     128  &  7.49  &  3.00  &     5.11  &  0.29  \\
             &   bg219g   &  8.668  &  2015-01-22   &  120  &     384  &  2.04  &  0.83  &     5.61  &  0.46  \\
             &   bg219i   &  2.292  &  2015-03-17   &  160  &     128  &  7.05  &  3.40  &  $-$4.43  &  0.28  \\
             &   bg219i   &  8.668  &  2015-03-17   &  240  &     384  &  1.93  &  0.92  &  $-$1.85  &  0.22  \\

\enddata
\tablecomments{Columns are as follows:
(1) J2000 name;
(2) Project codes of the observation;
(3) Observing frequency;
(4) the observation epoch;
(5) total integration time;
(6) total observing bandwidth;
(7) Major axis of the restoring beam (FWHM);
(8) Minor axis of the restoring beam (FWHM);
(9) Position angle of the major axis, measured from north through east;
(10) Off-source {\it rms} noise in the clean image;
}
\end{deluxetable}

\begin{deluxetable}{ccccccccc}
\centering
\tablecolumns{9}
\tabletypesize{\small}
\tablewidth{0pt}
\tablecaption{\label{model} Model fitting parameters}
\tablehead{
\colhead{J2000} & \colhead{Epoch}        & \colhead{Frequency} & \colhead{Comp.} & \colhead{$S_{int}$} & \colhead{R}       & \colhead{P.A.}       & \colhead{d}      & \colhead{$T_{\rm b}$} \\
\colhead{Name}  & \colhead {yyyy-mm-dd}  & \colhead {(GHz)}    & \colhead {}     & \colhead {(mJy)}     & \colhead{(mas)}   & \colhead{($\degr$)}  & \colhead{(mas)}  & \colhead{($10^{\rm 10}$K)}         \\
\colhead{(1)}   & \colhead{(2)}          & \colhead{(3)}       & \colhead{(4)}   & \colhead{(5)}       & \colhead{(6)}     & \colhead{(7)}        & \colhead{(8)}    & \colhead{(9)}          }
\startdata
J0906+4124 & 2012-04-22 & 8.355 & C  & 59.20$\pm$5.9  & ...           & ...       & 0.53$\pm$0.08  & 0.38  \\
           &            &       & W2 & 19.18$\pm$1.9  & 0.74$\pm$0.10 & $-$59.56  & 0.65$\pm$0.10  \\
           &            &       & E2 & 9.26$\pm$0.9   & 0.91$\pm$0.06 & 118.48    & 0.42$\pm$0.06  \\
           &            &       & W1 & 1.64$\pm$0.2   & 4.36$\pm$0.06 & $-$69.43  & 0.37$\pm$0.06  \\
           &            &       & E1 & 6.30$\pm$0.6   & 4.61$\pm$0.17 & 117.29    & 1.14$\pm$0.17  \\
J0909+1928 & 2006-01-28 & 4.845 & C  & 79.84$\pm$8.0  &  ...          & ...       & 0.18$\pm$0.03  & 13.03 \\
           &            &       & S2 & 8.16$\pm$0.8   & 4.00$\pm$0.16 & 165.14    & 1.04$\pm$0.16  \\
           &            &       & N1 & 4.81$\pm$0.5   & 4.63$\pm$0.15 & $-$3.71   & 1.02$\pm$0.15  \\
           &            &       & S1 & 11.71$\pm$1.2  & 8.26$\pm$0.40 & 140.36    & 2.67$\pm$0.40  \\
J0910+1841 & 2012-05-06 & 8.355 & C  & 20.81$\pm$2.1  &  ...          & ...       & 0.17$\pm$0.03  & 1.29  \\
J0933+1009 & 2012-01-13 & 8.355 & C  & 27.43$\pm$2.7  &  ...          & ...       & 0.32$\pm$0.06  & 0.47  \\
           &            &       & J1 & 6.49$\pm$0.6   & 1.44$\pm$0.04 & 99.20     & 0.24$\pm$0.04  \\
           & 2014-12-20 & 2.292 & C  & 47.68$\pm$4.8  &  ...          & ...       & 0.45$\pm$0.05  & 5.45  \\
           & 2014-12-20 & 8.668 & C  & 34.67$\pm$3.5  &  ...          & ...       & 0.15$\pm$0.02  & 2.48  \\
           & 2015-09-15 & 4.344 & C  & 39.21$\pm$0.4  &  ...          & ...       & 0.68$\pm$0.10  & 0.55  \\
           & 2015-09-15 & 7.624 & C  & 30.74$\pm$3.1  &  ...          & ...       & 0.24$\pm$0.04  & 1.14  \\
           &            &       & J1 & 5.52$\pm$0.6   & 1.31$\pm$0.02 & 101.43    & 0.10$\pm$0.02  \\
           & 2016-05-24 & 4.344 & C  & 38.21$\pm$3.8  & ...           & ...       & 0.77$\pm$0.15  & 0.42  \\
           & 2016-05-24 & 7.624 & C  & 18.45$\pm$1.8  & ...           & ...       & 0.22$\pm$0.03  & 0.79  \\
           &            &       & J1 & 2.62$\pm$0.3   & 1.32$\pm$0.03 & 102.33    & 0.54$\pm$0.08  \\
J0943+3614 & 2005-07-09 & 2.309 & C  & 134.11$\pm$13.2& ...           & ...       & 0.82$\pm$0.12  & 4.66  \\
           & 2005-07-09 & 8.646 & C  & 216.11$\pm$21.6& ...           & ...       & 0.42$\pm$0.06  & 2.04  \\
           & 2006-05-01 & 4.845 & C  & 205.17$\pm$20.5& ...           & ...       & 0.41$\pm$0.06  & 6.47  \\
           & 2012-05-06 & 8.646 & C  & 255.31$\pm$25.5& ...           & ...       & 0.44$\pm$0.07  & 2.20  \\
           & 2014-08-09 & 2.276 & C  & 225.71$\pm$22.5& ...           & ...       & 0.95$\pm$0.14  & 6.03  \\
           & 2014-08-09 & 8.646 & C  & 268.76$\pm$26.9& ...           & ...       & 0.45$\pm$0.07  & 2.21  \\
J1025+1022 & 2012-01-23 & 8.355 & C  & 26.18$\pm$2.6  &  ...          & ...       & 0.14$\pm$0.01  & 2.43  \\
           &            &       & W3 & 36.87$\pm$3.7  & 0.52$\pm$0.16 & $-$118.33 & 1.08$\pm$0.16  \\
           &            &       & W2 & 5.44$\pm$0.5   & 4.19$\pm$0.15 & $-$52.51  & 1.02$\pm$0.15  \\
           &            &       & E2 & 15.56$\pm$1.6  & 4.63$\pm$0.21 &   135.28  & 1.39$\pm$0.21  \\
           &            &       & W1 & 1.04$\pm$0.1   & 6.27$\pm$0.03 & $-$54.96  & 0.21$\pm$0.03  \\
           &            &       & E1 & 1.83$\pm$0.2   & 7.95$\pm$0.15 &  124.41   & 0.98$\pm$0.15  \\
J1037+4335 & 2011-01-20 & 8.355 & C  & 20.50$\pm$2.1  &  ...          & ...       & 0.25$\pm$0.04  & 0.61  \\
           &            &       & SW & 2.13$\pm$0.2   &41.42$\pm$0.10 & $-$142.63 & 0.66$\pm$0.10  \\
J1205+2031 & 2012-05-07 & 8.355 & C  & 22.72$\pm$2.3  &  ...          &  ...      & 0.07$\pm$0.01  & 8.41  \\
           &            &       & SW2& 6.65$\pm$0.7   & 0.95$\pm$0.04 & $-$123.84 & 0.29$\pm$0.04  \\
           &            &       & SW1& 1.59$\pm$0.2   & 2.54$\pm$0.01 & $-$127.10 & 0.04$\pm$0.01  \\
           &            &       & NE & 10.21$\pm$1.0  &14.93$\pm$1.01 &    49.41  & 6.72$\pm$1.01  \\
J1213+5044 & 2012-02-25 & 8.355 & C  & 29.71$\pm$3.0  &  ...          &  ...      & 0.47$\pm$0.07  & 0.24  \\
           &            &       & W  & 11.92$\pm$1.2  & 1.01$\pm$0.21 & $-$97.09  & 1.41$\pm$0.21  \\
           &            &       & E  &  5.03$\pm$0.5  & 1.61$\pm$0.22 &    82.45  & 1.45$\pm$0.22  \\
J1230+4700 & 2012-05-07 & 8.355 & C  & 18.57$\pm$1.9  &  ...          &  ...      & 0.26$\pm$0.04  & 0.51  \\
           &            &       & S2 & 13.82$\pm$1.4  & 0.89$\pm$0.06 &  174.79   & 0.43$\pm$0.06  \\
           &            &       & N  & 4.76$\pm$0.5   & 1.43$\pm$0.12 & $-$12.69  & 0.78$\pm$0.12  \\
           &            &       & S1 & 3.75$\pm$0.4   & 3.04$\pm$0.09 &  165.03   & 0.62$\pm$0.09  \\
J1246+1153 & 2012-06-03 & 8.355 & C  & 33.11$\pm$3.3  &  ...          &  ...      & 0.13$\pm$0.02  & 3.57  \\
           &            &       & J1 & 5.43$\pm$0.5   & 1.92$\pm$0.66 & $-$37.12  & 4.41$\pm$0.66  \\
J1559+2556 & 2010-12-12 & 8.640 & C  & 20.69$\pm$2.1  &  ...          &  ...      & 0.19$\pm$0.03  & 0.99  \\
           & 2011-02-07 & 8.355 & C  & 18.26$\pm$1.8  &  ...          &  ...      & 0.14$\pm$0.03  & 1.67  \\
           &            &       & J1 & 6.35$\pm$0.6   & 1.44$\pm$0.26 &     69.35 & 1.75$\pm$0.26  \\
           & 2013-07-08 & 4.344 & C  & 19.36$\pm$1.9  &  ...          &  ...      & 0.76$\pm$0.11  & 0.24  \\
           & 2013-07-08 & 7.624 & C  & 15.78$\pm$1.2  &  ...          &  ...      & 0.34$\pm$0.05  & 0.23  \\
           &            &       & J1 & 2.82$\pm$0.6   & 1.88$\pm$0.21 &     66.74 & 1.38$\pm$0.21  \\
J1604+1744 & 2006-04-03 & 4.845 & C  & 83.42$\pm$8.3  &  ...          &  ...      & 0.15$\pm$0.02  & 19.96 \\
           &            &       & J1 & 12.98$\pm$1.3  & 1.79$\pm$0.28 &  $-$171.71& 0.48$\pm$0.07  \\
           & 2012-01-08 & 8.355 & C  & 111.58$\pm$11.2&  ...          &  ...      & 0.19$\pm$0.03  & 4.78  \\
J1606+1814 & 2005-07-21 & 2.309 & C  & 131.72$\pm$13.2&  ...          &  ...      & 0.65$\pm$0.10  & 7.36  \\
           &            &       & J1 &  21.83$\pm$2.2 & 4.97$\pm$0.99 &  155.49   & 1.97$\pm$0.29  \\
           &            &       & J0 &   5.21$\pm$0.5 & 15.66$\pm$2.91&  156.98   & 0.54$\pm$0.08  \\
           & 2005-07-21 & 8.646 & C  & 99.42$\pm$10.0 &  ...          &  ...      & 0.09$\pm$0.03  & 20.87 \\
           &            &       & J3 & 23.90$\pm$2.4  & 0.92$\pm$0.01 &    165.63 & 0.06$\pm$0.01  \\
           &            &       & J2 & 17.81$\pm$1.8  & 2.30$\pm$0.09 &    161.12 & 0.61$\pm$0.09  \\
           &            &       & J1 & 6.72$\pm$0.7   & 5.39$\pm$0.08 &    158.51 & 0.55$\pm$0.08  \\
           & 2006-04-03 & 4.845 & C  & 140.42$\pm$14.0&  ...          &  ...      & 0.18$\pm$0.03  & 23.26 \\
           &            &       & J3 & 38.81$\pm$3.9  & 1.44$\pm$0.22 &  160.01   & 0.09$\pm$0.01  \\
           &            &       & J2 &  3.87$\pm$0.4  & 2.81$\pm$0.49 &  157.11   & 0.42$\pm$0.06  \\
           &            &       & J1 &  12.21$\pm$1.2 & 5.57$\pm$1.11 &  157.92   & 1.88$\pm$0.28  \\
           &            &       & J0 &   2.56$\pm$0.3 & 14.81$\pm$2.76&  158.68   & 2.48$\pm$0.37  \\
           & 2011-12-19 & 8.355 & C  & 95.61$\pm$9.6  &  ...          &  ...      & 0.30$\pm$0.05  & 1.91  \\
           &            &       & J3 & 15.01$\pm$1.5  & 1.74$\pm$0.10 &    157.57 & 0.67$\pm$0.10  \\
           &            &       & J1 & 4.93$\pm$0.5   & 5.71$\pm$0.15 &    158.44 & 1.01$\pm$0.15  \\
           & 2015-01-22 & 2.292 & C  & 114.78$\pm$11.5&  ...          &  ...      & 1.24$\pm$0.19  & 1.79  \\
           &            &       & J2 &  26.40$\pm$2.6 & 2.78$\pm$0.56 &  160.47   & 0.62$\pm$0.10  \\
           & 2015-01-22 & 8.668 & C  & 87.75$\pm$8.8  &  ...          &  ...      & 0.13$\pm$0.02  & 8.74  \\
           &            &       & J4 & 21.19$\pm$2.1  & 0.88$\pm$0.03 &    164.31 & 0.17$\pm$0.03  \\
           &            &       & J2 & 12.31$\pm$1.2  & 2.75$\pm$0.12 &    158.43 & 0.82$\pm$0.12  \\
           &            &       & J1 & 4.68$\pm$4.7   & 6.18$\pm$0.13 &    159.58 & 0.85$\pm$0.13  \\
           & 2015-03-17 & 2.292 & C  & 117.65$\pm$11.8&  ...          &  ...      & 1.23$\pm$0.18  & 1.88 \\
           &            &       & J2 &  36.34$\pm$3.6 & 2.72$\pm$0.54 &   160.01  & 1.47$\pm$0.22  \\
           &            &       & J1 &  12.44$\pm$1.2 & 6.48$\pm$1.26 &    155.64 & 1.81$\pm$0.27  \\
           &            &       & J0 &  6.64$\pm$0.7  & 15.44$\pm$3.08&    155.67 & 2.63$\pm$0.39  \\
           & 2015-03-17 & 8.668 & C  & 76.38$\pm$7.6  &  ...          &  ...      & 0.15$\pm$0.02  & 5.69  \\
           &            &       & J5 & 45.51$\pm$4.6  & 0.58$\pm$0.01 &    161.06 & 0.09$\pm$0.01  \\
           &            &       & J3 & 16.35$\pm$1.6  & 1.85$\pm$0.06 &    161.23 & 0.39$\pm$0.06  \\
           &            &       & J2 & 4.98$\pm$0.5   & 2.98$\pm$0.10 &    160.06 & 0.64$\pm$0.10  \\
           &            &       & J1 & 10.45$\pm$1.0  & 5.91$\pm$0.12 &    160.24 & 0.80$\pm$0.12  \\
\enddata
\tablecomments{Columns are as follows:
(1) J2000 name;
(2) Observation date;
(3) Frequency;
(4) Component;
(5) flux density of each component;
(6) Radius from the core;
(7) Position angle with respect to core, measured from north through east;
(8) size of gaussian model;
(9) core brightness temperature.
}

\end{deluxetable}

\begin{deluxetable}{ccccc}
\centering
\tablecolumns{6}
\tabletypesize{\small}
\tablewidth{0pt}
\tablecaption{\label{ratio of VLBI flux}Derived Parameters of four sources.}
\tablehead{
\colhead{J2000}
& \colhead{$S^{\rm VLA}_{\rm 1.4GHz}$}
& \colhead{$S^{\rm VLBI}_{\rm 1.4GHz}$}
& \colhead{$\alpha$}
& \colhead{R}
\\
\colhead{Name}
& \colhead {(Jy)}
& \colhead {(Jy)}
& \colhead {}
& \colhead{}  \\
\colhead{(1)}   & \colhead{(2)}       & \colhead{(3)}            & \colhead{(4)}             & \colhead{(5)}                      }
\startdata
J0933+1009 & 56.6  & 158.1   & $-$1.28 & 2.79    \\
J0943+3614 & 75.1  & 214.3   & 0.13    & 2.85    \\
J1559+2556 & 144.7 & 91.5    & $-$1.12 & 0.63    \\
J1606+1814 & 396.0 & 132.7   & $-$0.13 & 0.33    \\
\enddata
\tablecomments{Columns are as follows:
(1) J2000 name;
(2) NVSS 1.4 GHz flux density from \citet{1998AJ....115.1693C};
(3) total VLBI flux density at 1.4 GHz;
(4) spectral index;
(5) ratio of VLBI flux density to VLA flux density;
}
\end{deluxetable}

\end{document}